\documentclass[preprint,10pt,twocolumn]{emulateapj}
\usepackage{graphicx,amssymb,natbib}
\usepackage{amsmath}
\usepackage[top=2.5in, left=1in, right=0.5in, bottom=0in, paperwidth=8.5in, paperheight=11in]{geometry}

\newcommand{\about}{$\sim$}
\newcommand{\kms}{km s$^{-1}$}
\newcommand{\etal}{et al.}

\newcommand{\sqdeg}{deg$^{2}$}

\newcommand{\degrees}{\ensuremath{^\circ}}
\newcommand{\degree}{\degrees}

\newcommand{\bmagcutoff}{-16}

\def\be{\begin{equation}}
\def\ee{\end{equation}}
\shortauthors{Hallenbeck \etal}

\begin{document}
\shorttitle{Gas-Bearing Early-Type Dwarf Galaxies in Virgo}
\title{Gas-Bearing Early-Type Dwarf Galaxies in Virgo: Evidence for Recent Accretion}
\author{Gregory Hallenbeck, Emmanouil Papastergis, Shan Huang, Martha P. Haynes, Riccardo Giovanelli\altaffilmark{1}, Alessandro Boselli, Samuel Boissier, Sebastien Heinis\altaffilmark{2}, Luca Cortese\altaffilmark{3}, Silvia Fabello\altaffilmark{4}}
\altaffiltext{1}{Center for Radiophysics and Space Research, Space Sciences Building, Cornell University, Ithaca, NY 14853.
        {\textit{e-mail:}} ghallenbeck@astro.cornell.edu, papastergis@astro.cornell.edu, shan@astro.cornell.edu,
        haynes@astro.cornell.edu, riccardo@astro.cornell.edu}
\altaffiltext{2}{Laboratoire d'Astrophysique de Marseille - LAM, Universit\'e d'Aix-Marseille \& CNRS, UMR7326, 38 rue F. Joliot-Curie, 13388 Marseille Cedex 13, France {\textit{e-mail:}} Alessandro.Boselli@oamp.fr, Samuel.Boissier@oamp.fr, Sebastien.Heinis@oamp.fr}
\altaffiltext{3}{European Southern Observatory, Karl-Schwarzschild Str. 2, 85748 Garching bei Muenchen, Germany. {\textit{e-mail:}} lcortese@eso.org}
\altaffiltext{4}{Max-Planck-Institute f\"ur Astrophysik, D-85741 Garching, Germany. {\textit{e-mail:}} fabello@mpa-garching.mpg.de}
\begin{abstract}
We investigate the dwarf ($M_B>\bmagcutoff$) galaxies in the Virgo cluster in the radio, optical, and ultraviolet regimes.  Of the 365 galaxies in this sample, 80 have been detected in H \textsc{i} by the Arecibo Legacy Fast ALFA survey.  These detections include 12 early-type dwarfs which have H \textsc{i} and stellar masses similar to the cluster dwarf irregulars and BCDs.  In this sample of 12, half have star-formation properties similar to late type dwarfs, while the other half are quiescent like typical early-type dwarfs.  We also discuss three possible mechanisms for their evolution: that they are infalling field galaxies that have been or are currently being evolved by the cluster, that they are stripped objects whose gas is recycled, and that the observed H \textsc{i} has been recently reaccreted.  Evolution by the cluster adequately explains the star-forming half of the sample, but the quiescent class of early-type dwarfs is most consistent with having recently reaccreted their gas.
\end{abstract}

\section{Introduction}
Dwarf ellipticals (dEs) are rare in the field and in galaxy groups, but are the most common galaxies in clusters (\citealt{Binggeli1985}; \citealt{Caldwell1987};  \citealt{Trentham2009}). An ongoing question is whether cluster dEs form in situ, or whether the cluster environment strips infalling late-type galaxies of their gas, transforming them into dEs (\citealt{TullyShaya1984}; \citealt{Boselli2006}; \citealt{Boselli2008}). Driving such evolution are such processes as ram-pressure stripping \citep{Gunn1972}, starvation \citep{Larson1980}, and galaxy harassment \citep{Lin1983}. There is significant evidence that at least some subset of the dwarf elliptical population were once late-type galaxies. In the most luminous of early-type dwarfs (ETDs), late-type features such as stellar disks and faint spiral arms can be observed \citep{LiskerDisk}. Furthermore, rotational support or even rotational flattening is observed in a significant fraction of ETDs (\citealt{vanZee2004}; \citealt{Beasley2009}; \citealt{Toloba2011a}), and such galaxies share a Tully-Fisher plane with the dwarf irregulars. Rotationally supported ETDs are also observed to have younger stellar populations and are most likely to be found in the cluster outskirts \citep{Toloba2009}. 

The Virgo cluster is a rich ($>$1000 members), relatively nearby ($d\sim$17 Mpc) cluster. There is no clear single core, but rather two: a compact subcluster around M49, and a more massive and extended subcluster around M87. There are also several near-background `clouds' which are falling onto the cluster. Both its three dimensional structure (\citealt{Mei2007}) and velocity dispersion (\citealt{Binggeli1993}; \citealt{Drinkwater2001}; \citealt{Conselice2001}) suggest that the Virgo cluster is dynamically young. Due to its proximity, Virgo has been the target of many surveys, and all of its members have been targeted observationally. The most recent comprehensive compilation of Virgo member data can be found in GOLDMine (\citealt{GOLDMine}; \citealt{Gavazzi2005}), a multiwavelength aggregate of many data sources. Because it is so close, Virgo is also optimal for sensitive H \textsc{i} observations of both the dwarf galaxies and more massive members. It is the first cluster in which H \textsc{i} deficiency, a measure of how much less H \textsc{i} a galaxy contains compared to a field sample of similar optical properties, has been measured \citep{Davies1973}. Targeted H \textsc{i} observations exist for all massive late-type galaxies (\citealt{Gavazzi2005}), bright dwarf irregulars (\citealt{Hoffman1987}), and a significant number of dwarf ellipticals (\citealt{MHRG1986}; \citealt{Huchtmeier1986}; \citealt{vanDriel2000}; \citealt{Conselice2003}). Moreover, blind H \textsc{i} surveys have covered large fractions of Virgo, such as the H \textsc{i} Jodrell All Sky Survey (HIJASS; \citealt{Davies2004}) and the Arecibo Legacy Fast ALFA survey (ALFALFA; \citealt{ALFALFA1}). Recently, the VLA Imaging of Virgo in Atomic Gas (VIVA; \citealt{Chung2009}) survey has produced high-resolution H \textsc{i} maps of late-type galaxies in Virgo; this has allowed extensive study of stripping as it takes place (\citealt{Chung2007}; \citealt{Vollmer2008}). Direct H \textsc{i} evidence of galaxy harassment has been found by ALFALFA, which observed a $\sim$250 kpc tail off of NGC 4254 \citep{Haynes2007}.

The focus of this article is the dwarf galaxy population in the Virgo cluster, and specifically how dwarf galaxies form and evolve in the cluster environment, primarily through the study of their H \textsc{i} content. For this purpose, we will make extensive use of ALFALFA\footnote{The Arecibo Observatory is operated by SRI International under a cooperative agreement with the National Science Foundation (AST-1100968), and in alliance with Ana G. Méndez-Universidad Metropolitana, and the Universities Space Research Association.}, a blind H \textsc{i} survey which, in its current data release covering 40\% of the final survey area ($\alpha.40$; \citealp{ALFA40}), covers the Virgo cluster at declinations between 4\degrees\ and 16\degrees. Examining the H \textsc{i} content of Virgo has always been a goal of ALFALFA (\citealt{ALFALFANVirgo}; \citealt{Kent2008}), with the first global study performed by \citet{Gavazzi2008}, who investigated the scaling relations between the H \textsc{i} and luminous properties of ALFALFA-detected galaxies in the cluster. More recently, colors and star-formation properties using $\alpha.40$ and H$\alpha$ observations were described by \citet{Gavazzi2011a} and \citet{Gavazzi2011b}.

If dwarf ellipticals are the products of evolved dwarf irregulars and low luminosity spirals, then the efficient stripping of H \textsc{i} precedes quenched star formation and morphological transformation. In fact,  dwarf irregulars in Virgo have detectable reservoirs of H \textsc{i}, and the bulk of dwarf ellipticals are relatively gas-free, having low gas fraction (M$_\text{HI}$/M$_*$). Nonetheless, a small fraction of ALFALFA detections are elliptical galaxies. Early-type H \textsc{i}-detected galaxies included in previous ALFALFA data releases have been studied both in Virgo \citep{Alighieri2008} and in the field \citep{Grossi2009}. They found that very few Virgo ETGs had H \textsc{i}, and those that did were either massive galaxies which may be accreting from a companion, or dwarfs with peculiar morphologies. In the field, they found that a surprising 44\% of massive early-type galaxies were detected, possibly the results of major mergers, but only 13\% of dwarf ellipticals were detected---both much higher than the 2\% detection rate in Virgo.

With the near complete coverage of Virgo made available by $\alpha.40$, we compile a sample of 12 low-luminosity (M$_\text{B}>-16$) ETDs which are detected in H \textsc{i} by ALFALFA; this sample represents an almost two-fold increase from samples based on earlier ALFALFA catalogs. We also probe beneath the detection limit of ALFALFA using spectral stacking methods. In addition to their H \textsc{i} properties, we investigate their stellar populations and star formation activity using optical and ultraviolet photometry from the seventh data release of the Sloan Digital Sky Survey (SDSS DR7; \citealt{Abazajian2009}) and the GALEX satellite. An overview of the ALFALFA dwarf population, selected by H \textsc{i} mass, is presented in \citet{Huang2011}. An important result is that the within the ALFALFA dwarf sample, Virgo cluster members have lower gas fractions at a given M$_\text{HI}$ with a wide spread in the distributions of specific star formation rate (SFR/M$_*$).

Lastly, we consider possible evolutionary paths that would lead to the existence of H \textsc{i}-bearing dwarf galaxies with early-type morphologies in a cluster environment. The scenarios considered include the possibility of intrinsically gas-rich objects that have recently infallen onto the Virgo cluster, stripped objects with newly recycled H \textsc{i} from stars near the end of their lifetimes, and recent accreation of gas by formerly gas-free galaxies.

The paper is organized as follows: In section \ref{sec:datasample} we present the dwarf galaxy sample, and describe the selection process, as well as the H \textsc{i}, optical and UV data extraction process.  Section \ref{sec:results} describes the H \textsc{i} content, colors and star formation of the sample and the reference samples, and we discuss differences among dwarf galaxies with different morphologies.  In section \ref{sec:discussion} we present possible evolutionary paths for our sample of H \textsc{i}-detected ETDs, and argue about their consistency with our data.

\section{Data and Sample Selection}
\label{sec:datasample}

Our main sample is drawn from the Virgo Cluster Catalog (VCC) \citep{Binggeli1985}, with updated morphologies and assignments found in \citet{Binggeli1993}. We consider only VCC galaxies which are spectroscopically confirmed members of Virgo, and whose blue absolute magnitude in the VCC is fainter than M$_B=\bmagcutoff$ (corresponding to $m_B\gtrsim15$ at the distance of Virgo). Furthermore, we exclude galaxies that belong to one of the near-background complexes which appear to be infalling on Virgo, such as the W and W$^\prime$ clouds (\citealt{deVaucouleurs1961}), M cloud \citep{Ftaclas1984} and the southern extension \citep{Tully1982}. This results in a main sample composed of 365 Virgo members.  

Galaxies are then subdivided into classes according to their VCC morphology, which is based on high quality B-band photographic material.  Unless specified otherwise, we use the term `early-type dwarf' (ETD) to refer to galaxies classified as dwarf ellipticals (dE) and dwarf lenticulars (dS0), and the term `late-type dwarf' (LTD) to refer to galaxies classified as Magellanic irregulars (Im) and blue compact dwarfs (BCD) by the VCC.  The term `other dwarfs' refers to galaxies that do not belong to the previous two classes, generally classified as faint spirals, peculiar galaxies, or simply `?' by the VCC.\footnote{VCC 2062 was classified as a dwarf elliptical by \citeauthor{Binggeli1985}, but work by \citet{Duc2007} suggests that it is a tidal dwarf. As a result, we place it in the class of `other dwarfs.'} Overall, 275 out of the 365 galaxies in our main sample belong to the ETD class, 65 to the LTD class and 25 to the `other' class.

\subsection{H \textsc{I} Characteristics}
\label{sec:sample}

We use data from the ALFALFA blind H \textsc{i} survey, to get a statistically complete census of the atomic hydrogen content of galaxies in our main sample, specifically the $\alpha$.40 catalog which covers the Virgo cluster at declinations $4^\circ<\delta<16^\circ$, with a sensitivity of M$_{HI} \approx 10^{7.4}$ M$_\odot$ for a typical dwarf galaxy at the Virgo distance \citep[Section 6]{ALFA40}. This data release supercedes and expands the earlier ALFALFA releases (\citealt{ALFALFANVirgo}; \citealt{Saintonge2008}; \citealt{Kent2008}), which included Virgo coverage only between $8^\circ<\delta<16^\circ$. Table \ref{tab:cmpsample} lists the number of galaxies detected by ALFALFA.  In total, 80 galaxies in the main sample are included in $\alpha$.40: 51 LTDs (80\% of the class), 17 `other' (70\% of the class) and only 12 ETDs (5\% of the class). We divide these H \textsc{i}-detected ETDs into two subclasses, one blue and one red, based upon their SDSS $g-r$ color (see \S\ref{sec:photometry} for details on colors). This unusual subsample of 12 H \textsc{i}-detected dwarf galaxies with early-type morphologies is the main focus of the present paper. We present their H \textsc{i}, optical, UV, and fitted quantities (see \S\ref{sec:photometry}-\S\ref{sec:derived}) in Table \ref{tab:dwarfalfa}.

\begin{figure}
\begin{center}
\epsscale{0.75}
\plotone{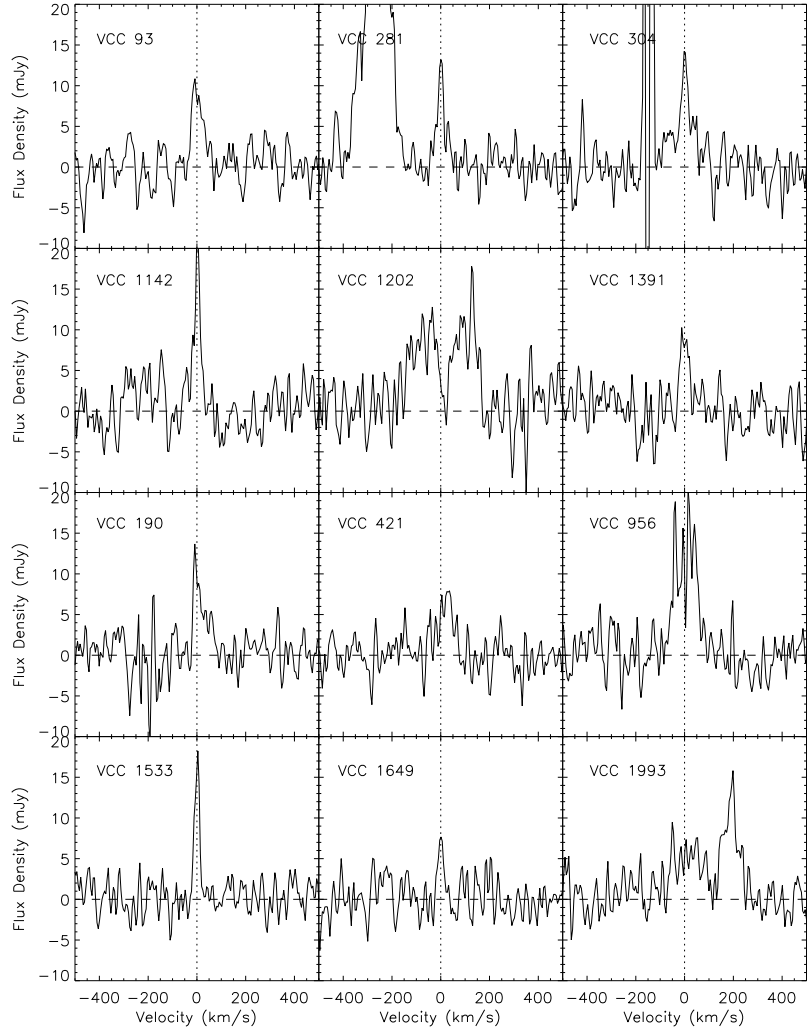}
\end{center}
\caption{The spectra of the 12 H \textsc{i}-detected ETDs, split into a blue subclass (top 6) and red subclass (bottom 6) based on SDSS the $g-r$ colors.  Horizontal dashed lines indicate the 0 mJy flux density level after baseline subtraction; vertical dotted lines indicate the line center velocity as reported in $\alpha.40$.
\label{fig:allspectra}
}
\end{figure}

Since the subsample of interest is comprised from such a limited number of objects, it is important to consider the reliability of these ETD H \textsc{i} detections individually. The H \textsc{i} spectra of all 12 galaxies can be seen in Figure \ref{fig:allspectra}, where the top 6 spectra are the blue subclass, the bottom 6 the red subclass).  The horizontal dashed line indicates the zero flux density level after baseline subtraction, and the vertical dotted line indicates the H \textsc{i} line center (see also Table \ref{tab:dwarfalfa} for their H \textsc{i} properties). All galaxies are well-separated from nearby, more massive H \textsc{i} emission, and their detections cannot be attributed to sidelobe contamination.

All but one of the $\alpha.40$ H \textsc{i} detections with S/N$<10$ (Table \ref{tab:dwarfalfa}) have been reobserved using more sensitive pointed observations, centered on the optical positions using Arecibo's L-band wide single pixel receiver. Of those, only VCC 421 was not confirmed. However, the H \textsc{i} emission of VCC 421 has a large ($1.1^\prime$) angular offset from the optical galaxy, and our failure to confirm it may merely be due to the LBW observing strategy of pointing at the optical galaxy, rather than centroided H \textsc{i} emission. Likewise, the emission of VCC 1649 has been confirmed to be offset (by $1.4^\prime$) from the optical galaxy, and so the association of the H \textsc{i} emission with this galaxy is uncertain. The ALFALFA spectrum of VCC 1202 is poor and very wide (W50$286$ km s$^{-1}$) because it is both near M87 and has only half the typical integration time for ALFALFA. The follow-up observations have better baselines. Of the 5 galaxies not yet re-observed, all are classified as `Code 1' sources in $\alpha$.40, that is they are confidently detected sources ($S/N_\text{HI} > 6.5$) with well-defined spectral profiles, and so are not of concern.

Only about half of the 12 H \textsc{i}-detected ETDs have previously reported H \textsc{i} observations in the literature. In particular, seven (VCC 93, 304, 1142, 1202, 421, 956, 1993) are included in the sample of H \textsc{i}-detected early-type Virgo galaxies of \citet{Alighieri2008}, which was based on earlier ALFALFA data releases. The H \textsc{i} sample of \citet{Gavazzi2008}, also based on earlier ALFALFA releases, includes two ETD detections (VCC 93, 304). Targeted H \textsc{i} observations (\citealt{Huchtmeier1986}; \citealt{Burstein1987}; \citealt{Duprie1996}) have reported 4 more gas-bearing dEs in Virgo; however none of them is included in our sample since they are all brighter than the $M_B>-16$ threshold. On the other hand, neither of the dwarf ellipticals (VCC 390 and 713) detected by the deep 21cm observations of \citet{Conselice2003} using Arecibo were detected by ALFALFA, despite having reported H \textsc{i} masses well above the ALFALFA completeness limit.
The Arecibo Galaxy Environment Survey (AGES; \citealt{AGES}) observations of the Virgo cluster \citep{AGESVirgo} detected 7 ETDs over 20 \sqdeg\, of which three appear in our sample (VCC 190, 611, and 1142); the other four do not pass our selection criteria. Of the three, only VCC 611 was not directly detected by ALFALFA (but see \S\ref{sec:gas}). Lastly, 6 of the 12 H \textsc{i}-detected ETD galaxies (VCC 93, 281, 304, 1142, 1533, 1649) are included in the complete sample of low H \textsc{i}-mass (log M$_\text{HI}$/M$_\odot < 7.7$) ALFALFA dwarfs of \citet{Huang2011}.

\subsection{Optical and Ultraviolet Photometry}
\label{sec:photometry}

In order to derive global stellar properties for the 12 H \textsc{i}-detected ETDs (stellar masses and star formation rates), we use optical images from the 7th data release of the Sloan Digital Sky Survey (SDSS DR7; \citealt{Abazajian2009}) and ultraviolet (UV) data from the GALEX mission. Figure \ref{fig:gallery} contains inverted SDSS images for all 12 galaxies; the images are 1.5$^\prime$ on a side (7.3 kpc at $d=16.7$ Mpc) and combine the $i$, $r$, and $g$ SDSS bands into the RGB color channels. Five galaxies (VCC 190, 956, 1142, 1391, 1533) are evidently very low surface brightness objects and their SDSS pipeline photometry was deemed `problematic.' To remedy this, the SDSS images of all 12 H \textsc{i}-detected ellipticals were manually reduced and their photometry was obtained individually with the use of IRAF/STSDAS tasks. Briefly, the corrected frames (\texttt{drC} files) in the $u,g,r,i,z$ bands were first smoothed to $2.5^{\prime\prime}$ resolution, and then background subtracted. After masking contaminating background and foreground sources, we fit elliptical isophotes to the $r$-band image of each galaxy. The derived apertures were applied to all other bands, and  Petrosian magnitudes based on these elliptical apertures was calculated, separately for each band.\footnote{The process described here is very similar, but not identical, to the SDSS pipeline measurement of Petrosian magnitudes (as described in \texttt{http://www.sdss.org/DR7/algorithms/photometry.html}). The main difference consists in the fact that the SDSS pipeline extracts measurements in circular rather than elliptical apertures. However, since the galaxies under consideration have generally low ellipticity, the difference in methodology is inconsequential for this study.} 

Ultraviolet photometry for the 12 H \textsc{i}-detected ETDs was obtained from the publicly available GALEX General Release 6 (GR6) database. Eight galaxies have deep ($t_{exp} \gtrsim 1000$ s) GALEX coverage, while the remaining four galaxies (VCC 281, 1391, 421, 1993) have only shallower coverage through the GALEX All-Sky Imaging Survey (AIS; $t_{exp} \approx 100$ s) images. Three galaxies (VCC 421, 1649, 1993) are not detected in the FUV band; in these cases the FUV flux measured at the position of the NUV detection by the GALEX pipeline (\texttt{fuv\_ncat\_flux}) was adopted as an upper limit on the galactic FUV flux. All magnitudes in Table \ref{tab:dwarfalfa} are corrected only for galactic extinction, based on the dust maps of \citet{Schlegel1998}. We do calculate internal extinction (see \S\ref{sec:derived}), but these corrections have high fractional uncertainty and are generally small (0.2-0.4 magnitudes) for these galaxies. In order to compare with similar works (\citealt{LiskerColors}; \citealt{Kim2010}), we do not include these extinction corrections in our reported magnitudes and colors. 
  
\subsection{Derived Quantities}
\label{sec:derived}

Atomic hydrogen (H \textsc{i}) masses are taken directly from $\alpha.40$. Distances to Virgo galaxies are based upon individualized subcluster assignments. We adopt a distance modulus of 31.11 for galaxies belonging to the A subcluster (M87 group), 31.08 for those belonging to the B subcluster (M49 group), and a weighted average of 31.09 for galaxies with uncertain subcluster assignment, according to the surface brightness fluctuation work of \citep{Mei2007}. We use primary distance measurements from the literature where available.

We estimate stellar masses and star formation rates (SFRs) for our galaxies using all available optical and UV bands following the method of \citet{Salim2007}, as described in \citet{Huang2011}. The computation involves fitting a library of model spectral energy distributions (SEDs) to the 5 SDSS and (when available) the 2 GALEX magnitudes. Full details of the method, as well as comparisons with other common estimators \citep[e.g.][]{Bell2003}, can be found in \citet{Huang2011}. Here we summarize the main points: model SEDs are generated by the \citet{Bruzual2003} stellar population synthesis code, assuming a \citet{Chabrier2003} stellar initial mass function. The star formation histories are a combination of continuous star formation and random bursts of star formation. We consider a large library of models with a broad range of internal extinctions ($0\leq \tau_V < 6$), $\mu$ factors (the fraction of optical depth that affects stellar populations older than 10 Myr; $0.1<\mu<1$), and metallicities ($0.1 Z_\odot \leq Z\leq 2 Z_\odot$). These parameter spaces are consistent with those of \citet{Salim2007}. The final physical properties are computed as the weighted average of all models, according to their fit likelihood. We stress that these extinction and metallicity values are not intended to encompass all possible values for a particular dwarf galaxy, but are expected to represent typical values for an ensemble of galaxies. Additionally, our stellar mass estimates are robust with respect to the exact value of the extinction parameter.

In cases where the FUV magnitude is not available, model SEDs are fit only to the 5 SDSS bands. We have verified that the derived stellar masses are not significantly affected by the exclusion of the UV photometry from the fitting process. On the other hand, SFRs derived from fitting just the optical bands tend to significantly overestimate the SFRs derived from fitting the full optical+UV SED (by up to $\approx 2$ dex, especially for galaxies with very low SFR). As a result, for galaxies not detected in the FUV we treat the SFR derived just from the optical data as an upper limit.    

\subsection{Reference samples}
\label{sec:reference}

\begin{figure}
\begin{center}
\epsscale{0.54}
\plotone{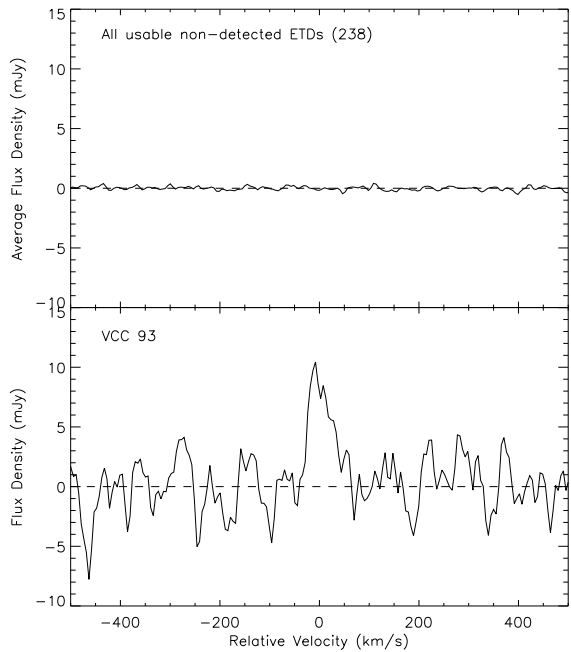}
\end{center}
\caption{
(top) Spectrum obtained by co-adding all useable 238 H \textsc{i} non-detected ETD spectra.  While the final noise level is greatly reduced, from \about2.2 mJy rms to 0.17 mJy, no H \textsc{i} emission signal is recovered. (bottom) Spectrum of VCC 93, for which the signal to noise of the spectrum extracted from the ALFALFA database is 6.  The x-axis values are velocities relative to the line center; as the stacked spectrum is produced by redshifting all component galaxies to 0 \kms, there is no true redshift for this subset.\label{fig:stackedspectrum}}
\end{figure}

In order to place the sample of 12 H \textsc{i}-detected ETDs in context, we analyze H \textsc{i}, optical and UV data for the rest of our main sample as well. This includes 68 H \textsc{i}-detected galaxies of late-type or `other' morphology and 285 H \textsc{i} non-detections, mostly of early-type morphology. We determine the average  gas mass of ETD non-detections by spectral stacking of the full ALFALFA datacubes. We follow the method of \citet{Fabello2011} and stack H \textsc{i} spectra using optical positions from the VCC and optical redshifts from the literature. In total 238 non-detected ETDs had acceptable spectra for stacking; galaxies whose emission would fall within the Galactic H \textsc{i} emission and galaxies that could be confused with more massive nearby sources were discarded. Figure \ref{fig:stackedspectrum} shows the spectrum produced by stacking all usable ETD spectra. The noise level in the stacked spectrum has clearly decreased with respect to the case of a single spectrum, but no H \textsc{i} emission is detected down to M$_\text{HI}/$M$_\odot = 10^{5.5}$. The greater area coverage of $\alpha.40$ allows for a significant improvement over the stacked mass sensitivity of M$_\text{HI}/$M$_\odot = 10^{6.1}$ achieved by the Virgo coverage of AGES \citep{AGESVirgo}.

We obtain optical and UV photometry for galaxies in the main sample through the SDSS DR7 and GALEX GR6 databases. The quality of the SDSS pipeline photometry of every galaxy was individually evaluated; galaxies that were `shredded' by the pipeline (i.e. assigned multiple photometric objects of comparable brightness) and low surface brightness objects with missing flux issues were flagged as `problematic' and excluded from further analysis. This quality cut affects mostly galaxies with M$_B \gtrsim -13.5$, of which about half have problematic photometry. A similar individual inspection of the quality of the GALEX pipeline photometry was performed for all galaxies in our main sample. Due to the lower resolution of the GALEX images ($\approx 5^{\prime\prime}$ seeing), problems related to shredding and missing flux of low surface brightness galaxies are far less common compared to SDSS. Overall, 314 galaxies have available NUV magnitudes, of which 240 were derived from deep GALEX images ($t_\text{exp}\gtrsim1000$ s) and 74 from shallower AIS images ($t_\text{exp}\approx100$ s). About 43\% of these galaxies were too faint to be detected in the FUV band; not surprisingly, all except 4 of the FUV non-detections correspond to early-type objects. All derived quantities for our main sample (HI mass, stellar mass, SFR etc.) are computed as described in \S\ref{sec:derived}. We note again that for galaxies that are not detected in the FUV, SFRs are based on fitting just the 5 SDSS bands and are used only as upper limits.

We also use the sample of low H \textsc{i}-mass ALFALFA dwarfs of \citet{Huang2011} as a comparison sample, representing the properties of gas-bearing field dwarfs. The H \textsc{i} masses, SDSS photometry and physical properties of the galaxies in the \citeauthor{Huang2011} sample are calculated as in this work (\S\ref{sec:photometry}-\S\ref{sec:derived}). However, UV magnitudes in \citet{Huang2011} are manually measured with the use of the GALPHOT image reduction package \citep{Haynes1999}, written in the IRAF/STSDAS environment. The difference between manually measured and pipeline magnitudes  is small in the NUV band ($\approx 0.20$ mag), while manually measured magnitudes tend to be brighter than pipeline values in the FUV band because the pipeline often misses low surface brightness flux (see Fig. 8 in \citealt{Huang2011}).

\section{Results}
\label{sec:results}
\subsection{Gas Content}
\label{sec:hidef}\label{sec:gas}

\begin{figure}
\begin{center}
\epsscale{0.8}
\plotone{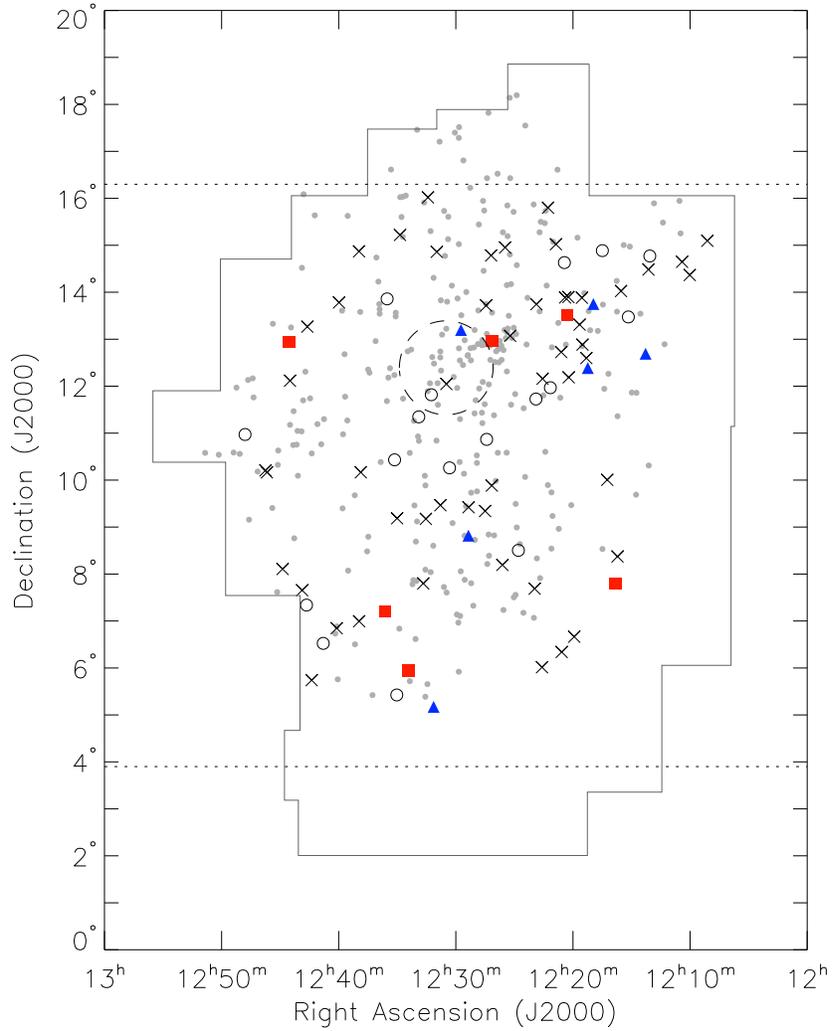}
\end{center}
\caption{Locations of the $\alpha$.40-detected dwarfs assigned membership in the Virgo cluster: xs are late-type dwarfs, open circles are `other dwarfs,' blue filled triangles are the blue subclass of ETDs and red filled squares are the red subclass of ETDs. H \textsc{i} non-detections with known redshifts are gray small filled circles. The dashed circle indicates a 1\degree\ ring around the location of M87.  Within this radius, the Strong (213 Jy; \citealt{Condon1998}) continuum emission of M87 causes an increase in system temperature and standing waves, therefore reducing the detectability of H \textsc{i} line sources (\citealt{ALFALFANVirgo}; \citealt{Kent2008}).  The horizontal dotted line indicates the Northern and Southern-most range of coverage of the  $\alpha$.40 survey \citep{ALFA40}.\label{fig:location}}
\end{figure}

\begin{figure}
\begin{center}
\epsscale{0.75}
\plotone{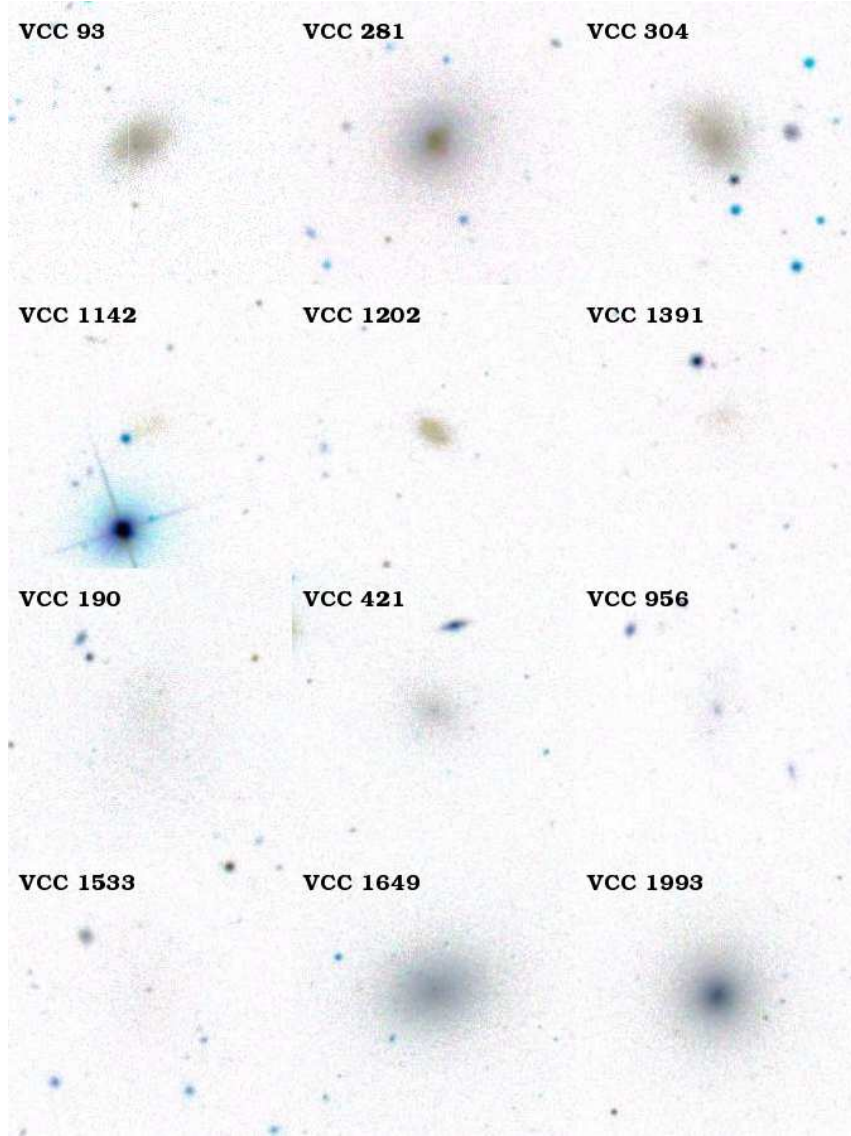}
\end{center}
\caption{Inverted SDSS images of all 12 VCC dwarf ellipticals detected by ALFALFA, 1.5$^\prime$ (7.2 kpc at D=16.7 Mpc) on a side.  The images combine the $g$,$r$, and $i$ SDSS bands (in the blue, green, and red color channels, respectively).  The top 6 belong to the blue ETD class (see \S\ref{sec:colors}), the bottom 6 the red ETD sample.  The surface brightness of VCC galaxies 190, 956, 1142, 1391 and 1533 are so low that the SDSS pipeline failed to yield suitable photometry. \label{fig:gallery}}
\end{figure}

It is well known that the Virgo cluster environment efficiently removes gas and advanced galaxy evolution. Even massive spirals are observed to be H \textsc{i} poor (\citealt{Davies1973}; \citealt{MHRG1986}), with small H \textsc{i} disks stripped inside the optical disk observed near the core, while relatively unperturbed disks are seen in the outskirts (\citealt{Cayatte1990}; \citealt{Chung2009}). Because of the removal of H \textsc{i}, the distribution of H \textsc{i} detections of dwarf galaxies in a survey like ALFALFA in Virgo is far from uniform. Figure \ref{fig:location} shows the sky distribution of the galaxies in our dwarf sample. Xs indicate the location of H \textsc{i} detected late-type dwarfs, open circles the H \textsc{i} detected `other' dwarfs, and small gray filled circles the non-detected dwarfs. The H \textsc{i} detected early-type dwarfs are plotted with different symbols based on subclass: the red ETDs are red filled squares, while the blue ETDs are blue filled triangles. Near M87, its strong continuum emission increases the system noise and sets up standing waves in the telescope optics and makes detection difficult; a 1\degree\ circle is drawn to indicate the region of contamination. Dashed lines indicate the northern and southern-most extent of the $\alpha.40$ survey. The outline corresponds to the plates of the original Virgo Cluster Survey \citep{Binggeli1985}. With the exception of two objects close to M87, the H \textsc{i} detected ETDs are found near the edges of the cluster along with the H \textsc{i} detected LTDs and `other' dwarfs.

\begin{figure}
\begin{center}
\epsscale{1.0}
\plotone{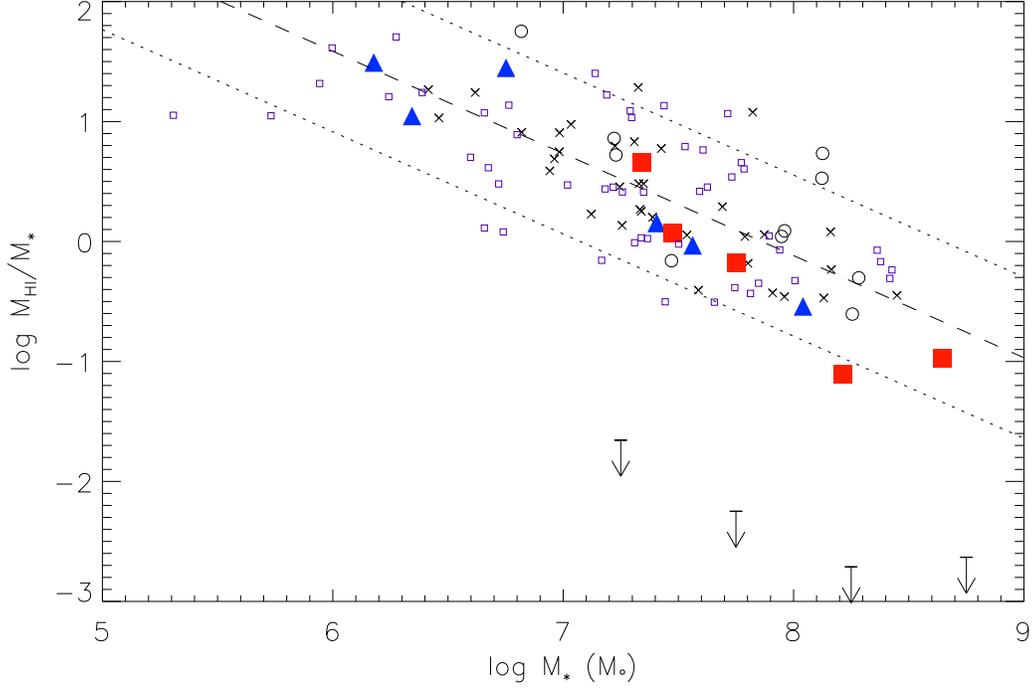}
\end{center}
\caption{Relationship between stellar mass and log gas fraction for the H \textsc{i}-detected sample. Symbols are the same as for Figure \ref{fig:location}. The lines show the linear trend and 2-$\sigma$ variation for the late-type galaxies only.  The arrows show 3$\sigma$ upper limits on the non-detections determined via stacking (see \S\ref{sec:sample}; \citealt{Fabello2011}). The small open purple squares are the ALFALFA dwarfs not in the Virgo cluster \citep{Huang2011}. As is typical for any population, the average gas fraction decreases with increasing stellar mass. It is unusual, however, that both the red and blue ellipticals appear H \textsc{i}-normal for late-type cluster galaxies of equivalent stellar mass.\label{fig:massmass}}
\end{figure}

Figure \ref{fig:massmass} shows the gas fraction as a function of SED-derived stellar mass for the H \textsc{i}-detected dwarfs.  The late-type dwarfs define a clear trend from high gas fraction at low stellar mass to low gas fraction at high stellar mass; a dashed line shows a linear fit to the LTDs with a dotted line showing a 2$\sigma$ variation around the line.  This distribution exhibits the same behavior observed by \citet{Gavazzi2011a} and \citet{Gavazzi2011b} for all Virgo galaxies; 
The H \textsc{i}-bearing dwarf galaxies in this sample are simply the low stellar mass end of all H \textsc{i}-bearing Virgo galaxies. Also plotted are the $\alpha.40$ dwarfs (small open purple squares; see \S\ref{sec:reference}), which fit into the same ranges as the LTDs, but with slightly more total scatter. The H \textsc{i}-stellar mass properties of the gas-bearing dwarfs in Virgo are thus similar to the field dwarf galaxies. At this level of inspection, there does not appear to be any significant difference between the two subclasses of H \textsc{i}-detected ETDs: both are well within the distribution of the late-type dwarfs at their respective stellar masses. The arrows in Figure \ref{fig:massmass} show the 5$\sigma$ upper limit of the non-detections as determined by spectral stacking. Even at this upper limit, the average non-detection is at least 2.5 dex poorer in H \textsc{i} gas fraction compared to the typical H \textsc{i} detected dwarf.

We visually inspected all 238 H \textsc{i} non-detected ETD spectra extracted for stacking (see \S\ref{sec:reference}) to look for any sign of gas just below ALFALFA's detection limit.  Of these, 9 galaxies had spectra with suggestive H \textsc{i} candidate signals of too low S/N to have been included in the $\alpha.40$ catalog. More sensitive pointed H \textsc{i} observations have been conducted of 8 of these 9 and confirmed the reality of one: VCC 611. The absence of a significant population just below the ALFALFA detection limit indicates that the H \textsc{i} detected ETDs are unlikely to be the tail of a wide distribution of H \textsc{i} masses, but are a separate population from the average very gas-poor ETDs.

\subsection{Colors and Star formation}
\label{sec:colors}\label{sec:sfr}

\begin{figure}
\begin{center}
\epsscale{0.59}
\plotone{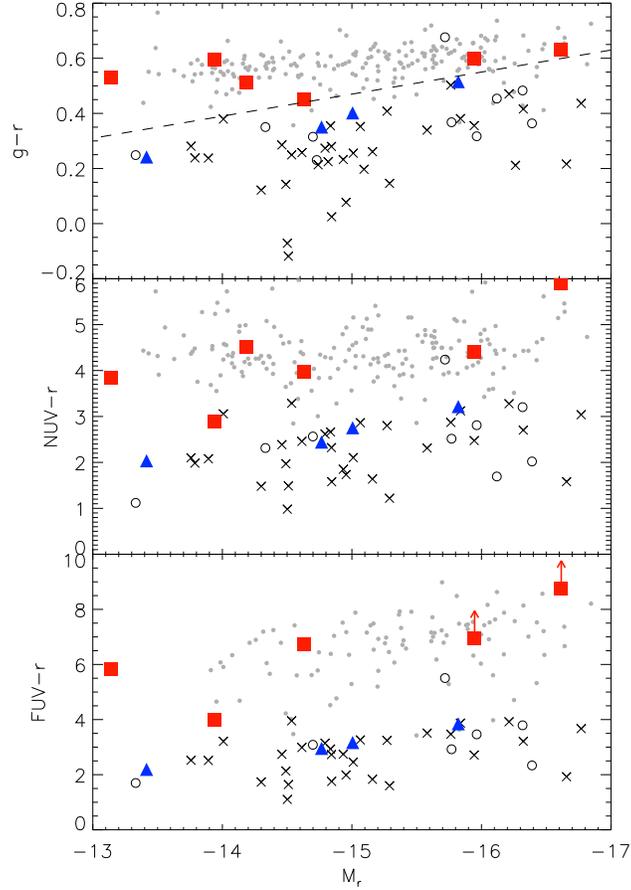}
\caption{Optical (SDSS) and UV (GALEX) colors for the dwarf galaxy sample for which acceptable photometry is available. Symbols are the same as for figure \ref{fig:location}. The dashed line shows our chosen split between the red and blue subclass; galaxies where $g-r< -0.08M_r - 0.73$ are considered blue. Two of the blue ETDs (VCC are so faint that they have $M_r\sim-11$; they are left unplotted as there are no other nearby galaxies for comparison. Arrows indicate galaxies for which the NUV$-r$ and FUV$-r$ colors are lower limits.\label{fig:colors}}
\end{center}
\end{figure}

\citet{LiskerColors} have shown that the optical colors of Virgo dwarf ellipticals follow a relationship similar to that of the more massive ellipticals: the fainter galaxies are in the $r$-band, the bluer their colors tend to be.  The same trend towards redder colors at brighter magnitudes was observed by \citet{Kim2010} in the NUV$-r$ and FUV$-r$ colors. In the FUV, these trends break down for the more massive ellipticals; this reddening is halted in ellipticals by the ultraviolet upturn (\citealt{Boselli2005}; \citealt{Boselli2008}; \citealt{Boselli2008other}). \citet{Kim2010} also found that both the late and early-types outside the cluster core are well separated in NUV$-r$ colors from the inner ellipticals; this separation is even greater in the FUV$-r$.  They argued that the bluer UV colors of galaxies in the cluster outskirts could be ascribed to higher star formation rates and higher gas fraction.  Figure \ref{fig:colors} shows the $g-r$, NUV$-r$, and FUV$-r$ colors plotted as a function of absolute $r$ band magnitude, with the same ranges plotted as \citet{Kim2010} Figure 1. The 12 gas-bearing ETDs as a class do not fit cleanly into either the red sequence or blue cloud: half lie within the red sequence, while the other half lie in the blue cloud.  For convenience, we have split the H \textsc{i} detected ETDs into two subclasses, galaxies that appear to belong to the blue cloud and galaxies that appear to belong to the red sequence. We find that the line of $g-r=-0.08M_r-0.73$ (plotted as a dashed line in Figure \ref{fig:colors}) adequately separates them, with the two faintest dwarfs in $r$-band (VCC 1142 and 1391) chosen to be blue. Two of the detected ETDs (VCC 190 and 281) are very close to the line, but clearly belong to the red sequence and blue cloud, respectively, when examining the NUV$-r$ and FUV$-r$ colors.

\begin{figure}
\begin{center}
\epsscale{0.7}
\plotone{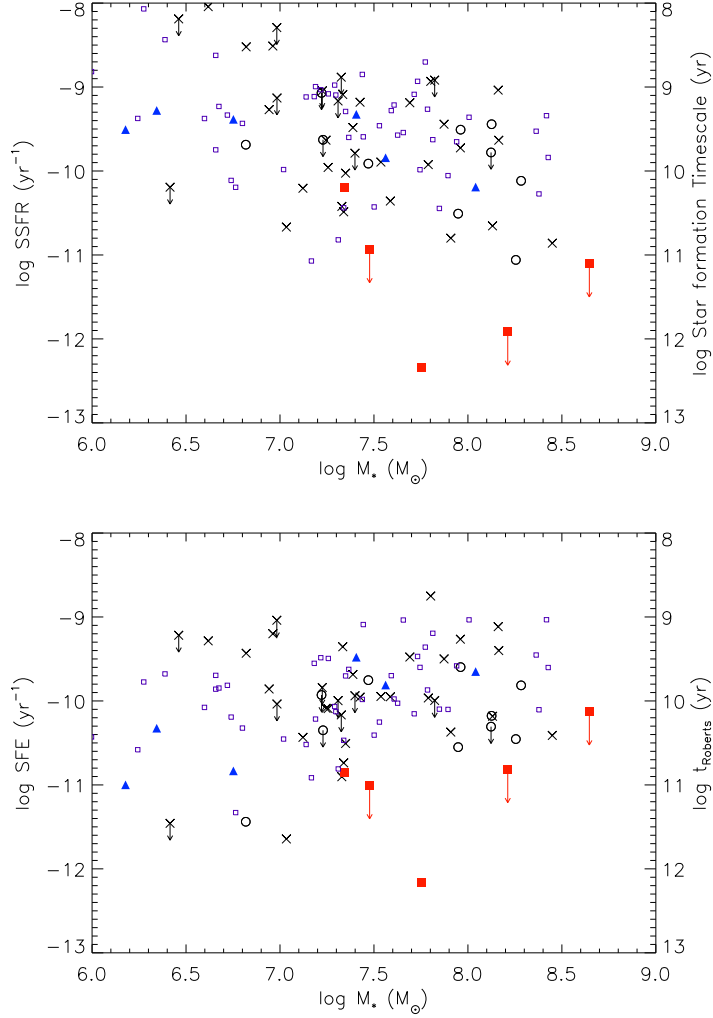}
\end{center}
\caption{Specific star formation rates (SFR/$M_*$; top) and star formation efficiencies (SFR/$M_\text{HI}$; bottom ) for the H \textsc{i}-detected dwarf galaxies as a function of stellar mass.  Symbols are identical to Figure \ref{fig:location} and arrows indicate upper limits. The small open purple squares are the $\alpha.40$ dwarfs not in the Virgo cluster. While there is large spread in SFR at the lower stellar masses, at high stellar mass (M$_*/$M$_\odot\gtrsim10^{7.5}$), the spread decreases and the detected late-type dwarfs, `other dwarfs' have somewhat elevated SFR compared to the red class of ETDs, while the galaxies in the blue class of ETDs fit into the lower half of the LTD SFR.\label{fig:SFR}}
\end{figure}

As blue color is a commonly used, albeit crude, indicator of star formation, we investigate star formation rates to begin understanding the difference in the red and blue subclass of the ETDs.  Figure \ref{fig:SFR} shows the trends of specific star formation rate (SFR$/$M$_*$) and star formation efficiency (SFR$/$M$_\text{HI}$) for all the H \textsc{i}-detected dwarfs as functions of stellar mass. The LTDs and `other' dwarfs form a band with width of \about2 dex in $\log$ yr$^{-1}$; the blue ETDs are also found in this region. At very low stellar masses, the SSFR and SFE have a much wider spread; this may be due to bursty, rather than continuous star formation history, as observed for the overall dwarf population by \citet{Huang2011}. The $\alpha.40$ field dwarfs (open purple squares; \citealt{Huang2011}) have a similar distribution in stellar mass,  SSFR, and SFE as the LTDs and blue ETDs.  With one exception, the red ETDs have low SSFR and SFE compared to the LTDs and blue ETDs of similar stellar mass. VCC 956, though a red galaxy, has a SSFR comparable to the bluer galaxies. The other red ETDs lie at the bottom of the blue galaxies' distribution or a dex lower.

\section{Evolutionary Hypotheses}
\label{sec:discussion}

We turn our attention to the possible histories of the H \textsc{i}-detected early-type dwarfs.  What mechanisms could allow a population of gas-bearing dwarf ellipticals to exist in a cluster environment today?  In addition to this question, an evolutionary history must also allow for the existence of two different subclasses of ETDs, one forming stars as strongly as the dwarf irregulars, and the other almost totally quiescent.  In the following, we discuss the merits of three possible evolutionary hypotheses for this particular population.  The first is that they are galaxies which were initially gas rich dwarf spirals or irregulars which are currently being stripped of their gas.  The second is that the gas originates from massive stars returning their hydrogen to the interstellar medium. The third scenario is that they are ellipticals which were previously red and `dead,' but have recently accreted a fresh supply of H \textsc{i}.


\subsection{Galaxies Being Stripped}

One evolutionary scenario is that ETDs are late-type galaxies, typically dwarf irregulars and low luminosity spirals, which have been evolved by the cluster. A variety of mechanisms have been proposed for the evolution from late-type galaxy to ETD in cluster. A strong contender is ram-pressure stripping \citep{Gunn1972}, where a galaxy moving through a cluster has the gas in its outskirts removed by the hot intra-cluster medium (ICM). The effect in the Virgo cluster is apparent enough that it has been observed directly in spirals at projected distances 0.6 Mpc $<d<$ 1 Mpc (\citealt{Chung2007}; \citealt{Vollmer2008}). The long-term consequence of ram-pressure stripping is a quenching of star formation, leading to a red, quiescent galaxy. \citet{Boselli2008} found that if a galaxy is in a first-pass highly eccentric orbit through the Virgo cluster center, this change can occur in as little as 100 Myr, much shorter than a cluster crossing time. In contrast, significant changes in color and star formation rate take longer, on the order of a Gyr. The dwarfs with the lowest stellar masses, including our sample, are expected to be the most susceptible to the evolving effects of the cluster. The ongoing ram-pressure stripping scenario may thus explain the gas-rich, star-forming ETDs we observe. However, for the H \textsc{i}-detected red ETDs, this explanation does not work: these galaxies have very low star formation rates, even though their H \textsc{i} has not yet been removed.

In another cluster evolution mechanism, galaxy harassment (\citealt{Moore1996}), a galaxy is perturbed by multiple close encounters with its neighbors.  \citet{Smith2010} found in simulations that, for dwarf galaxies infalling on Virgo, harassment significantly contributes to galaxy evolution in a relatively small number of cases ($<$ 15\%). However, in those cases, morphological transformation was strong, producing short-lived spiral arms in initially smooth disks, and signficant concentration of gas and star formation in the central part of the galaxy. \citet{LiskerBC} found a population of early-type dwarfs with disk-like features and star formation occuring in their central regions, and galaxy harassment may be the best explanation for these.  Of our H \textsc{i}-detected sample only VCC 281 exhibits this `blue center' quality; for this one object galaxy harassment is also a viable explanation for why it has evolved while maintaining a significant H \textsc{i} reservoir. However, ram-pressure stripping can also produce such a blue color gradient \citep{Boselli2008}.

\subsection{Recycled Gas}
\label{sec:recycle}

While ram-pressure stripping will efficiently remove gas from a dwarf galaxy after a single pass through the cluster, small quantities of H \textsc{i} can be replenished internally. As stars evolve, they return metal-enriched H \textsc{i} to the interstellar medium. We now consider the possible recycled origin of the gas, following the analysis of \citet{Boselli2008}. For an efficient ram-pressure stripping model, the amount of recycling can be quite significant. After a pass through the cluster core, star formation can then continue, at a star formation rate 1-2 dex below that found in a similar mass non-stripped dwarf, by feeding off of this recycled gas. This quenching of star formation will effectively move such a galaxy on the color-magnitude relation in FUV$-r$ from the blue cloud to the red sequence. We observe the red ETDs to have signficantly lowered SSFRs (by factors of 1-2 dex compared to the blue ETDs and LTDs), and they also fit solidly into the red sequence. However, the H \textsc{i} masses predicted by recycling are too low. For a dwarf galaxy beginning with $10^8$ M$_\odot$ of H \textsc{i}, after significant ram pressure stripping and gas recycling, the final gas mass is $\lesssim10^7$ M$_\odot$, below the ALFALFA detection limit. Additionally, \citet{Boselli2008} predict that such galaxies should have significantly less H \textsc{i} than the relatively unstripped LTDs in Virgo of the same stellar mass, instead we observe similar masses of H \textsc{i} in both populations.

\subsection{Re-Accreted Gas}
\label{sec:reaccrete}

\begin{figure}
\begin{center}
 \epsscale{1.0}
\plotone{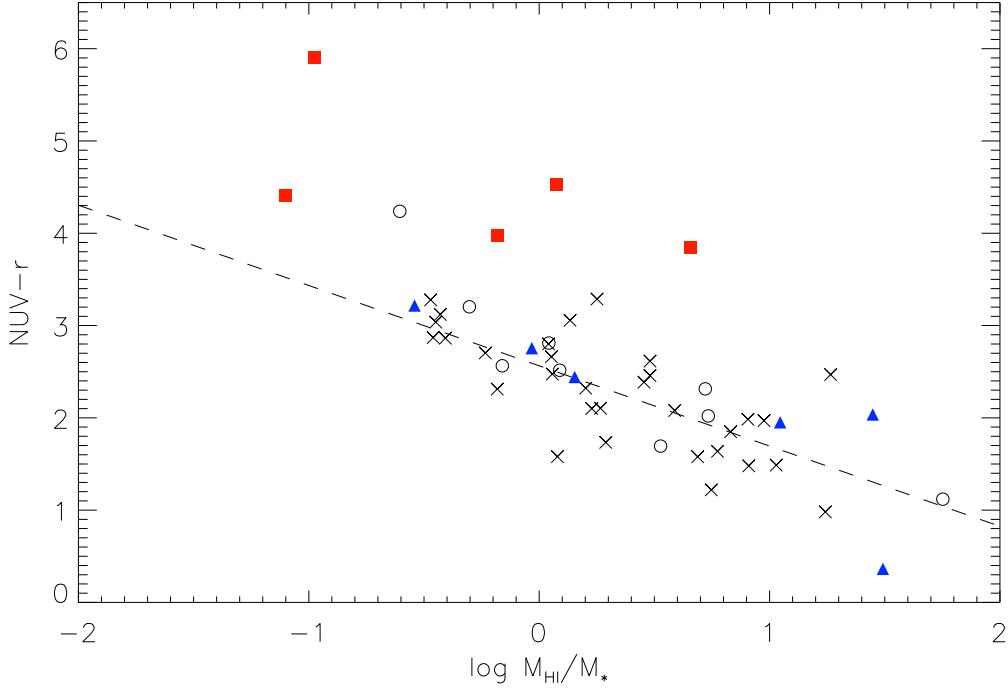}
\end{center}
\caption{
NUV$-r$ color as a function of H \textsc{i} gas fraction, with the same symbols as Figure \ref{fig:location}. The blue gas-bearing ETDs, LTDs, and `other' dwarfs form a clear line, with best-fit line plotted. The red ETDs have higher gas fraction than their NUV$-r$ color would indicate. \citet{Catinella2010} suggest that significant outliers, such as the red ETDs, are strong candidates for galaxies transitioning to or from the red sequence and blue cloud.
\label{fig:corteseplot}}
\end{figure}

\citet{Catinella2010} and \citet{Cortese2011} found a tight relationship between H \textsc{i} gas fraction, NUV$-r$ color, and stellar mass surface density, with NUV$-r$ being the best indicator of M$_\text{HI}/$M$_*$. Outliers are galaxies which are either forming too many or too few stars for their gas mass and are indicative of transition objects. More massive galaxies which are very red in NUV$-$H color which are otherwise H \textsc{i} normal show indication of recent mergers or gas accretion \citep{CorteseHughes2009}. Figure \ref{fig:corteseplot} plots the gas fraction as a function of NUV$-r$ color for all H \textsc{i} detections in Virgo. The blue ETDs, LTDs, and `other' dwarfs form a distinctive line with a best-fit line plotted. The red ETDs are all very strong outliers with high H \textsc{i} mass but low star formation: could they be reaccreting gas?

Dwarf galaxies in the Virgo cluster tend to be on highly eccentric, nearly free-fall orbits. The majority of such orbits will be spent relatively far from the cluster core, where neutral gas from the cosmic web will also be falling onto the cluster. At their furthest distance from the cluster, they will be moving at low velocity with respect to the cluster and distant gas, and while falling back into the cluster will have velocities similar to that of gas which is also free-falling onto the cluster. We can then follow the argument of \citet{Wei2010}, who examined a sample of low mass (M$_*<10^{10}$M$_\odot$) elliptical galaxies that appeared to be accreting gas in the field, and use the dynamical timescale to get a rough idea of the time needed to accrete gas. As H \textsc{i} in these sources is unresolved by ALFALFA (beam \about4'), we don't know the actual size of the gas clouds accreted. We choose half the Arecibo beam size at the distance of Virgo ($\sim$5 kpc) as an upper bound, and the optical sizes of the galaxies ($\sim$1.7 kpc) as a lower bound.  We assume that the total mass, accounting for dark matter content, is \about 10 times the baryonic content of the stars, H \textsc{i}, and Helium (\about30\% of the H \textsc{i}) together. We can then make an estimate of the gas infall time \citep{BinneyTremaine2008} as
\begin{equation} \label{eq:tinfall}
 t_\text{infall} \simeq \sqrt{\frac{\pi R^3}{ G \times 10 (M_*+1.3M_\text{gas})}}
\end{equation}
We get infall times for our sample that range from 24-470 Myr (see Table \ref{tab:dwarfalfa}). Given that most of the time of the galaxy's 1.5 Gyr orbit is spent at the outskirts of the cluster, these accretion timescales are all quite reasonable.

\begin{figure}
\begin{center}
 
\end{center}
\epsscale{0.8}
\plotone{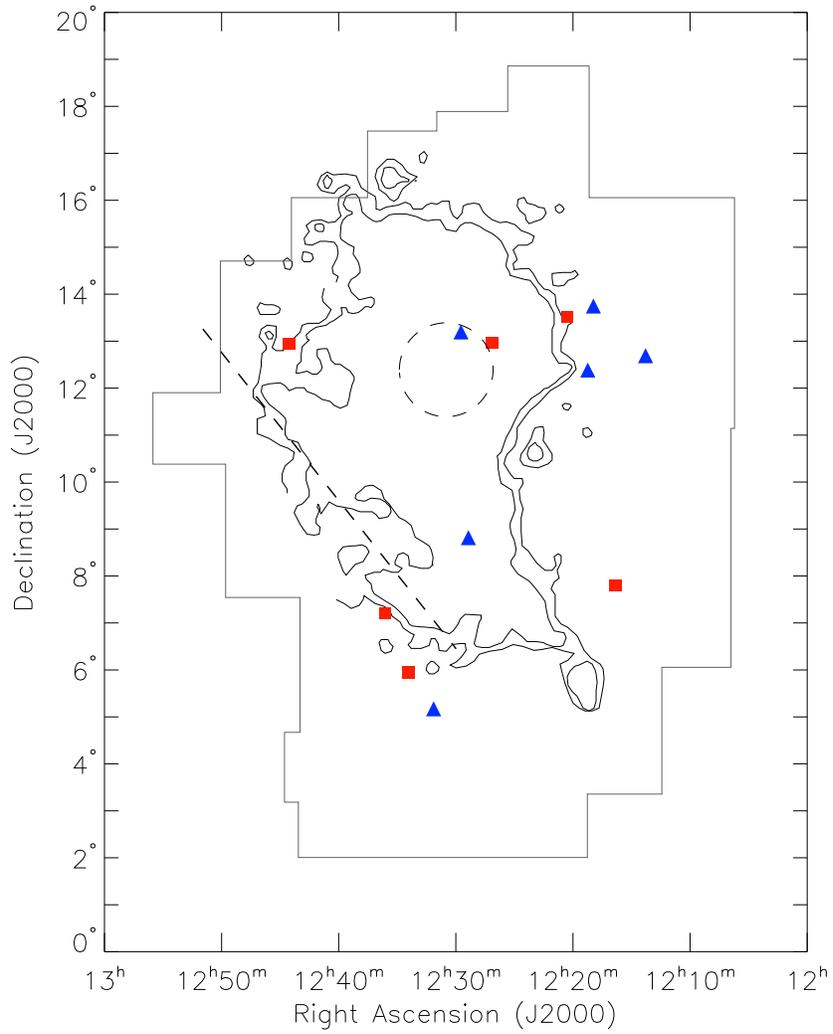}
\caption{Locations of the $\alpha$.40-detected early-type dwarfs. Blue triangles are the blue subclass, while red squares are the red subclass. Contours are the 3$\sigma$ and 5$\sigma$ detection limit of Rosat X-ray observations in the $0.4-2.4$ keV energy range \citep{Rosat1994}. Only contours associated with Virgo are plotted. The dark dashed line indicates onset of confusion with X-ray features of the Milky Way. The dashed circle indicates a 1\degree\ ring around M87, which is responsible for \about71\% of the X-ray emission.\label{fig:rosat}}
\end{figure}

But the Virgo ETDs galaxies are not isolated systems like those of \citet{Wei2010}, but rather are found in a cluster where the hot ICM will inhibit accretion and cooling of gas. Figure \ref{fig:rosat} shows the location of all H \textsc{i} detected ETDs in our sample with overlaid 3$\sigma$ and 5$\sigma$ X-ray contours (in the 0.4-2.4 keV range) from \citet{Rosat1994}. The dashed line indicates the onset of confusion with the north polar spur, an X-ray feature of the Milky Way, and only contours associated with Virgo are plotted. As with Figure \ref{fig:location}, the plates of \citet{Binggeli1985} and a 1\degree\ circle around M87 are plotted for reference. The 3$\sigma$ contour of the ROSAT map corresponds roughly to where the cluster hard ($0.4-3$ keV) X-ray flux is half of the background hard X-ray flux, with fluxes falling off as a power law \citep{Urban2011}. Of the 12 detections, 9 are at or clearly beyond this distance. Two of the exceptions, VCC 956 and 1202 are projected very close to M87 where the X-ray luminosity is very high; they may however be located far from the core, only projected onto M87. Likewise, VCC 1142's projected position puts it in a region of strong emission. For all three galaxies, we cannot be sure of how embedded they are in the ICM. For the other galaxies, we can make a rough estimate of how long it would take for the X-ray gas to evaporate accreting gas (\citealt{Cowie1977}; \citealt{Cowie1977b})
\begin{equation} \label{eq:tevap}
 t_\text{ev} \simeq 4.6\times 10^3\text{ yr} \left(\frac{M_\text{HI}}{M_\odot}\right)\left(\frac{R_\text{HI}}{\text{kpc}}\right)\left(\frac{T}{10^6\text{ K}}\right)^{-5/2}f_\text{mag}^{-1}
\end{equation}
Where R$_\text{HI}$ is the radius of the cloud (again, either 1.7 or 5 kpc), T is the temperature of the ICM medium, assuming an upper limit of $1.08$ keV at 3.9\degrees\ from M87 \citep{Urban2011}. $f_\text{mag}$ measures how the entanglement of the magnetic fields in the ICM quenches heat transfer; it is expected to be of order $\lesssim1/3$ \citep{Sarazin1986}. With the exception of the low H \textsc{i} mass galaxy VCC 1649, the estimated evaporation times are 121-721 Myr (see Table \ref{tab:dwarfalfa}), typically larger the dynamical infall times.	

So, the gas-bearing ETDs are generally outside radii from the cluster core where the cluster's X-ray emission dominates, their dynamical infall rates are significantly less than cluster crossing times and typically less than the lower-limit time for the X-ray medium to evaporate the cool gas, making recent accretion a definite possibility for all but four of the dwarf ETDs (the three inside significant X-ray emission, and the low H \textsc{i} mass VCC 1649). For the red subclass of gas-bearing ETDs, recent accretion is also the only evolutionary history of the three discussed which can adequately explain their optical, UV, and H \textsc{i} properties.

\section{Conclusions}

We examined the sample of all 365 VCC dwarf ($M_B > -16$) galaxies with known redshifts. While not as sensitive as observations targeting individual galaxies, ALFALFA's blind coverage gives us a statistically complete view of the atomic gas content of the Virgo cluster population, both for detected and non-detected galaxies. We use ALFALFA's current $\alpha.40$ \citep{ALFA40} data release, which expands the previous Virgo coverage (\citealt{ALFALFANVirgo}; \citealt{Kent2008}) to declinations $4^\circ<\delta<16^\circ$. Of the 365 dwarf galaxies, 80 are detected in H \textsc{i} in the $\alpha.40$ catalog. As is to be expected, a large fraction of these galaxies (68 out of 80) are late-types and peculiar galaxies, but 12 are classified as early-types by the VCC.

The gas-bearing population of dwarf galaxies show a clear tendency to avoid the regions of Virgo closest to M87 and M49. Despite being early-type galaxies, the H \textsc{i}-detected ETDs have H \textsc{i} gas fractions similar to the H \textsc{i}-detected LTDs of the same stellar mass. However, the upper limit on the gas fraction for the non-detected ETD population is approximately 2.5 dex lower than the H \textsc{i} detected dwarfs, and there does not appear to be a significant population bridging the gap between the gas bearing and the gas poor populations. 

In agreement with \citet{LiskerColors} and \citet{Kim2010}, we find that the dwarf ellipticals follow a tight color-magnitude relation with bluer $g-r$, NUV$-r$, and FUV$-r$ colors at fainter M$_r$. The H \textsc{i}-detected late-type dwarfs trace a very broad distribution which is well-separated from the red sequence in $g-r$ (0.4 magnitudes); the separation is larger in NUV$-r$ and FUV$-r$ (2 and 4 magnitudes, respectively). The H \textsc{i} detected ETDs are unusual in that they do not as a class sit wholly in either the red sequence or blue cloud: half are clearly red while the other half are blue. The blue gas-bearing ETDs have SSFRs and SFEs similar to the LTDs and the `other' dwarfs, while the red ETDs typically have rates a few dex lower than the blue subclass.

Lastly, we consider the various possible evolutionary histories of the bas-bearing ETDs. It is quite possible that the blue ETDs are on their first pass through the cluster, and are unstripped or being stripped. VCC 281 stands out from the other blue ETDs in that it has a `blue-center' suggestive of a disk galaxy evolved by galaxy harassment but is also consistent with ram pressure stripping. However, stripping cannot explain the red subclass of ETDs, which are gas-rich but not forming stars. Models of gas returned into the ISM by stars at the end of their lives \citep{Boselli2008} cannot account for the mass of H \textsc{i} observed in either the blue or red subclass. The gas-bearing red ETD subset is most consistent with having already traveled through the cluster core, been stripped, and now at lying the edge of the cluster where H \textsc{i} is being accreted from the cosmic web. All of the red ETDs, with the exception of VCC 956 ($\sim1^\circ$ from M87), are in a region of the cluster where the dominant hard X-ray flux arises from background sources, rather than the ICM. The dynamical timescales for these galaxies to re-accrete their gas are significantly less than one orbital time and typically larger than the time it would take the ICM to evaporate the newly accreted gas.
\\

\subsubsection*{Acknowledgements}
The authors would like to acknowledge the work of the entire ALFALFA collaboration team
in observing, flagging, and extracting the catalog of galaxies used in this work.
The ALFALFA team at Cornell is supported by NSF grants AST-0607007 and AST -1107390 and
by a grant from the Brinson Foundation.\\
\\
The research leading to these results has received funding from the European Community's Seventh Framework Programme (/FP7/2007-2013/) under grant agreement No 229517.\\
\\
GALEX is a NASA Small Explorer, launched in 2003 April operated for NASA by the California
 Institute of Technology under NASA contract NAS5-98034. We gratefully acknowledge NASA's
support for construction, operation and science  analysis for the GALEX mission, developed
in cooperation with the Centre National d'Etudes Spatiales of France and the Korean Ministry
of Science and Technology. EP, MPH and RG acknowledge support for this work
from the GALEX Guest Investigator program under NASA grant NNX10AI01G.\\
\\
Funding for the SDSS and SDSS-II has been provided by the Alfred P. Sloan Foundation,
the participating institutions, the National Science Foundation, the US Department
of Energy, the NASA, the Japanese Monbukagakusho, the Max Planck Society and
the Higher Education Funding Council for England. The SDSS Web Site is http://www.sdss.org/.
The SDSS is managed by the Astrophysical Research Consortium for the
participating institutions. The participating institutions are the American
Museum of Natural History, Astrophysical Institute Potsdam, University of Basel,
University of Cambridge, Case Western Reserve University, University of Chicago,
Drexel University, Fermilab, the Institute for Advanced Study, the Japan
Participation Group, Johns Hopkins University, the Joint Institute for Nuclear
Astrophysics, the Kavli Institute for Particle Astrophysics and Cosmology, the
 Korean Scientist Group, the Chinese Academy of Sciences (LAMOST), Los Alamos
National Laboratory, the Max Planck Institute for Astronomy, the MPA, New Mexico
State University, Ohio State University, University of Pittsburgh, University
of Portsmouth, Princeton University, the United States Naval Observatory and
the University of Washington.

\begin{centering}
\begin{deluxetable}{lrrrr}
\tablecolumns{5}
\tablewidth{0pt}
\tabletypesize{\scriptsize}
\tablecaption{Dwarf Sample Properties\label{tab:cmpsample}}
\tablehead{
  \colhead{Selection} & \colhead{Total} & \colhead{ETDs} & \colhead{LTDs} & \colhead{Other}
}
\startdata
BST Dwarfs     & 502 & 413 & 89 & --\\
\hline
$M_B<-16$      & 365 & 275 & 65 & 25\\
\hline
ALFALFA        &  80 &  12 & 51 & 17\\
SDSS           & 271 & 211 & 43 & 17\\
GALEX-NUV      & 252 & 197 & 50 &  5\\
GALEX-FUV      & 140 &  89 & 46 &  5\\
\hline
\enddata
\tablecomments{Total number of galaxies and number of galaxies by morphology in samples discussed in \S\ref{sec:datasample}.  BST Dwarfs are the galaxies identified as dwarfs in \citep{Binggeli1985} which have redshifts, while $M_B<-16$ consists of all VCC galaxies with redshifts and faint $M_B$. The ALFALFA, SDSS, and GALEX samples are are all direct subsets of the blue magnitude based sample.}
\end{deluxetable}
\end{centering}

\begin{centering}
\begin{deluxetable}{lrrlrrrrrrrrrrrrrr}
\tiny
\tablecolumns{17}
\setlength{\tabcolsep}{2pt}
\tabletypesize{\scriptsize}
\tablecaption{Dwarf Elliptical H \textsc{i} Detections\label{tab:dwarfalfa}}
\tablehead{
  \omit & \tiny  VCC & \tiny  AGC & \tiny  Morph. & \tiny $m_B$ & \tiny log M$_\text{HI}$ & \tiny  log M$_*$ & \tiny  SN & \tiny $W_{50}$ & \tiny  V$_\odot$ & \tiny log SFR & \tiny  M$_r$ & \tiny  g-r & \tiny  NUV-r & \tiny  FUV-r & \tiny  t$_\text{infall}$ & \tiny t$_\text{evap}$ \\
\omit  & \omit      & \omit        & \omit        & \omit     & \tiny M$_\odot$~~ & \tiny M$_\odot$~ & \omit     & \tiny \kms & \tiny\kms & \tiny ~M$_\odot$ yr$^{-1}$ & \omit & \omit & \omit & \omit & \tiny  Myr & \tiny Myr\\
\omit  & \tiny (1)~ & \tiny (2)~~~ & \tiny ~~~(3) & \tiny (4) & \tiny (5)         & \tiny (6)        & \tiny (7) & \tiny (8)  & \tiny (9) & \tiny (10) & \tiny (11) & \tiny (12) & \tiny (13) & \tiny (14) & \tiny (15) & \tiny (16)
}
\startdata
\tiny Blue    & \tiny   93 & \tiny 223286 & \tiny dE2     & \tiny 16.3 & \tiny 7.56 & \tiny 7.41 & \tiny  6.0 & \tiny  50 & \tiny  841 & \tiny    -1.92 & \tiny -14.8 & \tiny 0.35 & \tiny 2.44 & \tiny    2.96 & \tiny  $64-334$ & \tiny $180-560$\\
        & \tiny  281 & \tiny 220321 & \tiny dS0     & \tiny 15.3 & \tiny 7.50 & \tiny 8.04 & \tiny  5.4 & \tiny  32 & \tiny  248 & \tiny    -2.15 & \tiny -15.8 & \tiny 0.52 & \tiny 3.22 & \tiny    3.85 & \tiny  $42-220$ & \tiny $156-488$\\
        & \tiny  304 & \tiny 223407 & \tiny dE1pec? & \tiny 16.3 & \tiny 7.53 & \tiny 7.56 & \tiny  6.2 & \tiny  35 & \tiny  132 & \tiny    -2.28 & \tiny -15.0 & \tiny 0.40 & \tiny 2.76 & \tiny    3.16 & \tiny  $61-318$ & \tiny $167-523$\\
        & \tiny 1142 & \tiny 222021 & \tiny dE1     & \tiny 19.0 & \tiny 7.67 & \tiny 6.18 & \tiny 13.7 & \tiny  27 & \tiny 1306 & \tiny    -3.33 & \tiny -11.2 & \tiny 0.26 & \tiny 0.36 & \tiny    0.49 & \tiny  $70-360$ & \\
        & \tiny 1202 & \tiny 223724 & \tiny dE?     & \tiny 20.0 & \tiny 8.20 & \tiny 6.75 & \tiny  8.0 & \tiny 286 & \tiny 1215 & \tiny    -2.63 & \tiny -13.4 & \tiny 0.24 & \tiny 2.04 & \tiny    2.20 & \tiny  $38-195$ & \\
        & \tiny 1391 & \tiny 223819 & \tiny dE1     & \tiny 18.5 & \tiny 7.39 & \tiny 6.34 & \tiny  5.4 & \tiny  42 & \tiny 2308 & \tiny    -2.94 & \tiny -12.3 & \tiny 0.20 & \tiny 1.95 & \tiny    3.00 & \tiny  $94-488$ & \tiny $121-379$\\
\hline
\tiny Red     & \tiny  190 & \tiny 223355 & \tiny dE4     & \tiny 18.0 & \tiny 7.57 & \tiny 7.75 & \tiny 10.2 & \tiny  32 & \tiny 2352 & \tiny    -4.59 & \tiny -14.6 & \tiny 0.45 & \tiny 3.98 & \tiny    6.72 & \tiny  $54-283$ & \tiny $183-573$\\
        & \tiny  421 & \tiny 223445 & \tiny dE2     & \tiny 17.0 & \tiny 7.55 & \tiny 7.48 & \tiny  5.1 & \tiny 111 & \tiny 2098 & \tiny $<$-3.45 & \tiny -14.2 & \tiny 0.51 & \tiny 4.52 & \tiny     --- & \tiny  $62-325$ & \tiny $110-344$\\
        & \tiny  956 & \tiny 222857 & \tiny dE1,N:  & \tiny 18.8 & \tiny 8.00 & \tiny 7.34 & \tiny 10.3 & \tiny 102 & \tiny 2151 & \tiny    -2.85 & \tiny -13.1 & \tiny 0.53 & \tiny 3.84 & \tiny    5.82 & \tiny  $42-220$ & \\
        & \tiny 1533 & \tiny 223873 & \tiny dE2,N   & \tiny 18.0 & \tiny 7.44 & \tiny  *   & \tiny 10.5 & \tiny  21 & \tiny  648 & \tiny   *      & \tiny -13.9 & \tiny 0.59 & \tiny 2.89 & \tiny    3.98 & \tiny $<91-474$ & \tiny $136-425$\\
        & \tiny 1649 & \tiny 223913 & \tiny dE3,N:  & \tiny 15.7 & \tiny 7.11 & \tiny 8.21 & \tiny  3.3 & \tiny  28 & \tiny  972 & \tiny $<$-3.70 & \tiny -16.0 & \tiny 0.60 & \tiny 4.40 & \tiny $>$6.95 & \tiny  $41-212$ & \tiny $64-198$\\
        & \tiny 1993 & \tiny 220977 & \tiny E0      & \tiny 15.3 & \tiny 7.67 & \tiny 8.65 & \tiny  5.8 & \tiny 119 & \tiny  925 & \tiny $<$-2.45 & \tiny -16.6 & \tiny 0.63 & \tiny 5.90 & \tiny $>$8.76 & \tiny  $24-126$ & \tiny $231-722$\\
\enddata
\tablecomments{Properties of the dwarf ellipticals detected in H \textsc{i} by ALFALFA, split based on g-r color from the SDSS.  Subclass: The blue subclass is defined by lower g-r color as a function of absolute $r$-band magnitude; Column 1: VCC identifier (from \citealt{Binggeli1985}); Column 2: Arecibo General Catalog number \citep{ALFA40}; Column 3: Galaxy morphology (\citealt{Binggeli1985}; \citealt{Binggeli1993}); Column 4: VCC apparent B-band magnitude \citep{Binggeli1985}; Column 5: H \textsc{i} mass, from $\alpha.40$; Column 6: stellar mass as detailed in \S\ref{sec:derived}; Columns 7 through 9: H \textsc{i} detection signal to noise, H \textsc{i} line width at 50\% level, and heliocentric velocity, from $\alpha.40$; Column 10: star formation rate, as discussed in \S\ref{sec:derived}; Columns 11 through 14: absolute r-band magnitude, SDSS and GALEX colors, for galaxies where data exists.  Colors are corrected for galactic extinction only, using the SDSS pipeline correction; Column 12: Range of Estimated gas infall times based on Equation \ref{eq:tinfall}, assuming infall from the optical radius (lower bound) or half the size of Arecibo beam (upper bound); Column 13: Estimated gas evaporation time based on Equation \ref{eq:tevap}, assuming accreted gas cloud was half the size of the Arecibo beam (lower bound) or the optical radius (upper bound).\\
* The fit for VCC 1533 fails because the $z$-band photometry is bad: it is fainter in $z$-band than in other bands, while all of our models have $z$ band as the brightest band.}
\end{deluxetable}
\end{centering}


\begin{thebibliography}{}
\bibitem[Abazajian et al.(2009)]{Abazajian2009} Abazajian, K.~N., et al.\ 2009, \apjs, 182, 543 
\bibitem[Auld et al.(2006)]{AGES} Auld, R., et al.\ 2006, \mnras, 371, 1617
\bibitem[Beasley et al.(2009)]{Beasley2009} Beasley, M.~A., Cenarro, A.~J., Strader, J., \& Brodie, J.~P.\ 2009, \aj, 137, 5146 
\bibitem[Bell et al.(2003)]{Bell2003} Bell, E.~F., McIntosh, D.~H., Katz, N., \& Weinberg, M.~D.\ 2003, \apjs, 149, 289 
\bibitem[Binggeli et al.(1993)]{Binggeli1993} Binggeli, B., Popescu, C.~C., \& Tammann, G.~A.\ 1993, \aaps, 98, 275 
\bibitem[Binggeli et al.(1985)]{Binggeli1985} Binggeli, B., Sandage, A., \& Tammann, G.~A.\ 1985, \aj, 90, 1681 
\bibitem[Binney \& Tremaine(2008)]{BinneyTremaine2008} Binney, J., \& Tremaine, S.\ 2008, Galactic Dynamics: Second Edition, by James Binney and Scott Tremaine.~ISBN 978-0-691-13026-2 (HB).~Published by Princeton University Press, Princeton, NJ USA, 2008.,  
\bibitem[B{\"o}hringer et al.(1994)]{Rosat1994} B{\"o}hringer, H., Briel, U.~G., Schwarz, R.~A., Voges, W., \& Tr{\"u}mper, J.\ 1994, \nat, 368, 828 
\bibitem[Boselli et al.(2008a)]{Boselli2008other} Boselli, A., Boissier, S., Cortese, L., \& Gavazzi, G.\ 2008a, \aap, 489, 1015 
\bibitem[Boselli et al.(2008b)]{Boselli2008} Boselli, A., Boissier, S., Cortese, L., \& Gavazzi, G.\ 2008b, \apj, 674, 742 
\bibitem[Boselli \& Gavazzi(2006)]{Boselli2006} Boselli, A., \& Gavazzi, G.\ 2006, \pasp, 118, 517 
\bibitem[Boselli et al.(2005)]{Boselli2005} Boselli, et al.\ 2005, \apjl, 629, L29 
\bibitem[Bruzual \& Charlot(2003)]{Bruzual2003} Bruzual, G., \& Charlot, S.\ 2003, \mnras, 344, 1000 
\bibitem[Burstein et al.(1987)]{Burstein1987} Burstein, D., Krumm, N., \& Salpeter, E.~E.\ 1987, \aj, 94, 883 
\bibitem[Caldwell(1987)]{Caldwell1987} Caldwell, N.\ 1987, \aj, 94, 1116 
\bibitem[Catinella et al.(2010)]{Catinella2010} Catinella, B., et al.\ 2010, \mnras, 403, 683 
\bibitem[Cayatte et al.(1990)]{Cayatte1990} Cayatte, V., van Gorkom, J.~H., Balkowski, C., \& Kotanyi, C.\ 1990, \aj, 100, 604 
\bibitem[Chabrier(2003)]{Chabrier2003} Chabrier, G.\ 2003, \pasp, 115, 763 
\bibitem[Chung et al.(2009)]{Chung2009} Chung, A., van Gorkom, J.~H., Kenney, J.~D.~P., Crowl, H., \& Vollmer, B.\ 2009, \aj, 138, 1741 
\bibitem[Chung et al.(2007)]{Chung2007} Chung, A., van Gorkom, J.~H., Kenney, J.~D.~P., \& Vollmer, B.\ 2007, \apjl, 659, L115 
\bibitem[Condon et al.(1998)]{Condon1998} Condon, J.~J., Cotton, W.~D., Greisen, E.~W., Yin, Q.~F., Perley, R.~A., Taylor, G.~B., \& Broderick, J.~J.\ 1998, \aj, 115, 1693 
\bibitem[Conselice et al.(2001)]{Conselice2001} Conselice, C.~J., Gallagher, J.~S., III, \& Wyse, R.~F.~G.\ 2001, \apj, 559, 791 
\bibitem[Conselice et al.(2003)]{Conselice2003} Conselice, C.~J., O'Neil, K., Gallagher, J.~S., \& Wyse, R.~F.~G.\ 2003, \apj, 591, 167 
\bibitem[Cortese et al.(2011)]{Cortese2011} Cortese, L., Catinella, B., Boissier, S., Boselli, A., \& Heinis, S.\ 2011, \mnras, 415, 1797 
\bibitem[Cortese \& Hughes(2009)]{CorteseHughes2009} Cortese, L., \& Hughes, T.~M.\ 2009, \mnras, 400, 1225 
\bibitem[Cowie \& McKee(1977)]{Cowie1977} Cowie, L.~L., \& McKee, C.~F.\ 1977, \apj, 211, 135 
\bibitem[Cowie \& Songaila(1977)]{Cowie1977b} Cowie, L.~L., \& Songaila, A.\ 1977, \nat, 266, 501 
\bibitem[Davies \& Lewis(1973)]{Davies1973} Davies, R.~D., \& Lewis, B.~M.\ 1973, \mnras, 165, 231 
\bibitem[Davies et al.(2004)]{Davies2004} Davies, J., et al.\ 2004, \mnras, 349, 922 
\bibitem[de Vaucouleurs(1961)]{deVaucouleurs1961} de Vaucouleurs, G.\ 1961, \apjs, 6, 213 
\bibitem[di Serego Alighieri et al.(2007)]{Alighieri2008} di Serego Alighieri, et al.\ 2007, \aap, 474, 851 
\bibitem[Drinkwater et al.(2001)]{Drinkwater2001} Drinkwater, M.~J., Gregg, M.~D., \& Colless, M.\ 2001, \apjl, 548, L139 
\bibitem[Duc et al.(2007)]{Duc2007} Duc, P.-A., Braine, J., Lisenfeld, U., Brinks, E., \& Boquien, M.\ 2007, \aap, 475, 187 
\bibitem[Duprie \& Schneider(1996)]{Duprie1996} Duprie, K., \& Schneider, S.~E.\ 1996, \aj, 112, 937 
\bibitem[Fabello et al.(2011)]{Fabello2011} Fabello, S., Catinella, B., Giovanelli, R., Kauffmann, G., Haynes, M.~P., Heckman, T.~M., \& Schiminovich, D.\ 2011, \mnras, 411, 993 
\bibitem[Ftaclas et al.(1984)]{Ftaclas1984} Ftaclas, C., Struble, M.~F., \& Fanelli, M.~N.\ 1984, \apj, 282, 19 
\bibitem[Gavazzi et al.(2003)]{GOLDMine} Gavazzi, G., Boselli, A., Donati, A., Franzetti, P., \& Scodeggio, M.\ 2003, \aap, 400, 451 
\bibitem[Gavazzi et al.(2005)]{Gavazzi2005} Gavazzi, G., Boselli, A., van Driel, W., \& O'Neil, K.\ 2005, \aap, 429, 439 
\bibitem[Gavazzi et al.(2012a)]{Gavazzi2011a} Gavazzi, G., Fumagalli, M., Galardo, V., Grossetti, F., Boselli, A., Giovanelli, R., Haynes, M.~P., \& Fabello, S.\ 2012a, \aap, in submission
\bibitem[Gavazzi et al.(2012b)]{Gavazzi2011b} Gavazzi, G., Fumagalli, M., Galardo, V., Grossetti, F., Boselli, A., Giovanelli, R., Haynes, M.~P., \& Fabello, S.\ 2012b, \aap, in submission
\bibitem[Gavazzi et al.(2008)]{Gavazzi2008} Gavazzi, G., et al.\ 2008, \aap, 482, 43 
\bibitem[Giovanelli et al.(2005)]{ALFALFA1} Giovanelli, R., et al.\ 2005, \aj, 130, 2598 
\bibitem[Giovanelli et al.(2007)]{ALFALFANVirgo} Giovanelli, R., et al.\ 2007, \aj, 133, 2569 
\bibitem[Grossi et al.(2009)]{Grossi2009} Grossi, M., et al.\ 2009, \aap, 498, 407 
\bibitem[Gunn \& Gott(1972)]{Gunn1972} Gunn, J.~E., \& Gott, J.~R., III 1972, \apj, 176, 1 
\bibitem[Haynes \& Giovanelli(1986)]{MHRG1986} Haynes, M.~P., \& Giovanelli, R.\ 1986, \apj, 306, 466 
\bibitem[Haynes et al.(2007)]{Haynes2007} Haynes, M.~P., Giovanelli, R., \& Kent, B.~R.\ 2007, \apjl, 665, L19 
\bibitem[Haynes et al.(1999)]{Haynes1999} Haynes, M.~P., Giovanelli, R., Salzer, J.~J., Wegner, G., Freudling, W., da Costa, L.~N., Herter, T., \& Vogt, N.\ 1999, \aj, 117, 1668 
\bibitem[Haynes et al.(2011)]{ALFA40} Haynes, M.~P., et al.\ 2011, \aj, 142, 170 
\bibitem[Hoffman et al.(1987)]{Hoffman1987} Hoffman, G.~L., Glosson, J., Helou, G., Salpeter, E.~E., \& Sandage, A.\ 1987, \apjs, 63, 247 
\bibitem[Huang et al.(2012)]{Huang2011} Huang, S., Haynes, M.~P., Giovanelli, R., Brinchmann, J., Stierwalt, S., \& Neff, S.~G.\ 2012, \aj, in press
\bibitem[Huchtmeier \& Richter(1986)]{Huchtmeier1986} Huchtmeier, W.~K., \& Richter, O.-G.\ 1986, \aaps, 64, 111 
\bibitem[Kent et al.(2008)]{Kent2008} Kent, B.~R., et al.\ 2008, \aj, 136, 713 
\bibitem[Kim et al.(2010)]{Kim2010} Kim, S., Rey, S.-C., Lisker, T., \& Sohn, S.~T.\ 2010, \apjl, 721, L72 
\bibitem[Larson et al.(1980)]{Larson1980} Larson, R.~B., Tinsley, B.~M., \& Caldwell, C.~N.\ 1980, \apj, 237, 692 
\bibitem[Lin \& Faber(1983)]{Lin1983} Lin, D.~N.~C., \& Faber, S.~M.\ 1983, \apjl, 266, L21 
\bibitem[Lisker et al.(2006a)]{LiskerBC} Lisker, T., Glatt, K., Westera, P., \& Grebel, E.~K.\ 2006a, \aj, 132, 2432 
\bibitem[Lisker et al.(2006b)]{LiskerDisk} Lisker, T., Grebel, E.~K., \& Binggeli, B.\ 2006b, \aj, 132, 497 
\bibitem[Lisker et al.(2007)]{LiskerColors} Lisker, T., Grebel, E.~K., Binggeli, B., \& Glatt, K.\ 2007, \apj, 660, 1186 
\bibitem[Mei et al.(2007)]{Mei2007} Mei, S., et al.\ 2007, \apj, 655, 144 
\bibitem[Moore et al.(1996)]{Moore1996} Moore, B., Katz, N., Lake, G., Dressler, A., \& Oemler, A.\ 1996, \nat, 379, 613 
\bibitem[Paudel et al.(2010)]{Paudel2010} Paudel, S., Lisker, T., Kuntschner, H., Grebel, E.~K., \& Glatt, K.\ 2010, \mnras, 405, 800 
\bibitem[Saintonge et al.(2008)]{Saintonge2008} Saintonge, A., Giovanelli, R., Haynes, M.~P., Hoffman, G.~L., Kent, B.~R., Martin, A.~M., Stierwalt, S., \& Brosch, N.\ 2008, \aj, 135, 588 
\bibitem[Salim et al.(2007)]{Salim2007} Salim, S., Rich, R.~M., Charlot, S., et al.\ 2007, \apjs, 173, 267 
\bibitem[Sarazin(1986)]{Sarazin1986} Sarazin, C.~L.\ 1986, Reviews of Modern Physics, 58, 1 
\bibitem[Schlegel et al.(1998)]{Schlegel1998} Schlegel, D.~J., Finkbeiner, D.~P., \& Davis, M.\ 1998, \apj, 500, 525 
\bibitem[Smith et al.(2010)]{Smith2010} Smith, R., Davies, J.~I., \& Nelson, A.~H.\ 2010, \mnras, 405, 1723 
\bibitem[Taylor et al.(2012)]{AGESVirgo} Taylor, R., Davies, J.~I., Auld, R., \& Minchin, R.~F.\ 2012, arXiv:1203.3094
\bibitem[Toloba et al.(2011)]{Toloba2011a} Toloba, E., Boselli, A., Cenarro, A.~J., Peletier, R.~F., Gorgas, J., Gil de Paz, A., \& Mu{\~n}oz-Mateos, J.~C.\ 2011, \aap, 526, A114 
\bibitem[Toloba et al.(2009)]{Toloba2009} Toloba, E., et al.\ 2009, \apjl, 707, L17 
\bibitem[Trentham \& Tully(2009)]{Trentham2009} Trentham, N., \& Tully, R.~B.\ 2009, \mnras, 398, 722 
\bibitem[Tully(1982)]{Tully1982} Tully, R.~B.\ 1982, \apj, 257, 389 
\bibitem[Tully \& Shaya(1984)]{TullyShaya1984} Tully, R.~B., \& Shaya, E.~J.\ 1984, \apj, 281, 31 
\bibitem[Urban et al.(2011)]{Urban2011} Urban, O., Werner, N., Simionescu, A., Allen, S.~W., \& B{\"o}hringer, H.\ 2011, \mnras, 414, 2101 
\bibitem[van Driel et al.(2000)]{vanDriel2000} van Driel, W., Ragaigne, D., Boselli, A., Donas, J., \& Gavazzi, G.\ 2000, \aaps, 144, 463 
\bibitem[van Zee et al.(2004)]{vanZee2004} van Zee, L., Skillman, E.~D., \& Haynes, M.~P.\ 2004, \aj, 128, 121 
\bibitem[Vollmer et al.(2008)]{Vollmer2008} Vollmer, B., Soida, M., Chung, A., van Gorkom, J.~H., Otmianowska-Mazur, K., Beck, R., Urbanik, M., \& Kenney, J.~D.~P.\ 2008, \aap, 483, 89 
\bibitem[Wei et al.(2010)]{Wei2010} Wei, L.~H., Kannappan, S.~J., Vogel, S.~N., \& Baker, A.~J.\ 2010, \apj, 708, 841 
\end{thebibliography}
\end{document}